\documentclass[twocolumn,showpacs,preprintnumbers,amsmath,amssymb]{revtex4}

\usepackage{graphicx}
\usepackage{dcolumn}
\usepackage{bm}

\begin{document}


\title{GLLH EM Invisible Cloak With Novel Front Branching And \\
 Without Exceeding Light Speed Violation }

\author{Ganquan Xie}
 \altaffiliation[Also at ]{GL Geophysical Laboratory, USA, glhua@glgeo.com}
\author{Jianhua Li, Lee Xie, Feng Xie}%
 \email{GLGANQUAN@GLGEO.COM}
\affiliation{%
GL Geophysical Laboratory, USA
}%

\hfill\break

\date{\today}

\begin{abstract}
In this paper, for the first time in the world, we propose a new electromagnetic (EM) cloak without superluminal propagation and without time delay. Using Global and Local (GL) electromagnetic non-scattering modeling and inversion with a distinctive class of materials, $a_{\alpha \beta}\log ^\alpha (b_{\alpha \beta}/h) h^\beta$ (GLLH Cloak), the named GLLH invisible cloak is developed in $arXiv:1005.3999V1$ in 2010. After 16 years, this paper is version 2 and 3 of $arXiv:1005.3999V1$. The refractive index of the GLLH cloak material is greater than or equal to one. Spherical harmonic analysis and the GL modeling and inversion method are used to simulate the electromagnetic wave propagation through the GLLH cloak without superluminal effects and without time delay. At the time step $119dt$, 
the EM wave inside of  the GLLH  cloak propagates slower than light speed, moreover, its two crescent front peaks intersect at
a front branching point. The novel EM wave propagation and front branching in the GLLH cloak obtained by GL EM modeling are presented. The GLLH invisible cloak is not from coordinate transformation. Find an invisible cloak is the non scattering problem. Using Global and Local (GL) electromagnetic non-scattering modeling and inversion with a distinctive class of materials is one method in our paper in $arXiv:1005.3999V1$ in 2010. Another method is some 0 to R1 radial coordinate transformations with superluminal and infinite phase velocity. From the GLLH invisible cloak, we discovered positive space and negative space and invisible science. Using a new negative infinity to 0 quasi topological transformation, we discovered new isotropic electromagnetic invisible GLHUA sphere; anisotropic electromagnetic invisible GLHUA cloak.
In the GLLH cloak, the wave front is curved as a crescent like and without superluminal and time delay. Open question: Can we construct a 3D Kakeya set where line segments of length  represent the vector field of EM wave ray-tracing propagation through GLLH, GLHUA cloaks, and the GLHUA sphere? 
All copyright and patent of the GLLH EM cloaks and GL modeling
and inversion
methods are reserved by authors in GL Geophysical Laboratory.
\end{abstract}

\maketitle

\section{\label{sec:level1}INTRODUCTION} 
In a 2010 preprint on $arXiv:1005.3999$[1] and at PIERS in Moscow 2009[2], 2026 preprint on $https://papers.ssrn.com/sol3/papers.cfm?abstract_id=6062994$ , we proposed a new electromagnetic (EM) cloak without superluminal propagation and without time delay for the first time in the world. Using Global and Local (GL) electromagnetic non-scattering modeling and inversion with a distinctive class of materials, $a_{\alpha \beta}\log ^\alpha (b_{\alpha \beta}/h) h^\beta$  [3][4][5], the named GLLH invisible cloak without infinite speed violation, without superluminal propagation, and without time delay is proposed and developed in $arXiv:1005.3999V1$ in 2010.   After 16 years, this paper is version 2 and 3 of $arXiv:1005.3999V1$. The relative refractive index of the GLLH cloak material, is greater than one or equal to one. Spherical harmonic analysis and the GL modeling and inversion method are used to simulate and verified the electromagnetic wave propagation through the GLLH cloak without superluminal effects and without time delay. The novel EM wave propagation and front branching in the GLLH cloak obtained by GL EM modeling are presented. The wave front propagation in GLLH cloak is discontinuous. The GLLH invisible cloak materials absorb incident wave and create outgoing wave simultaneously. In 2001, we used Global Integral and Local Differential (GILD) EM modeling and inversion[6] to detect the fly model imaging from EM sources in some frequency band. A strange cloak phenomenon was discovered. The cloak protects the fly object from the exterior EM field detection. We called the strange phenomena as GILD effect and published it in paper [7].  For investigating the strange double layer clothes phenomena[7], we developed
Global and Local field electromagnetic modeling method [3] and GL Metro Carlo inversion method [5]. Using Global and Local (GL) electromagnetic non-scattering modeling and inversion with a distinctive class of materials, $a_{\alpha \beta}\log ^\alpha (b_{\alpha \beta}/h) h^\beta$ [3][4][5], the GLLH invisible cloak without infinite speed violation, without superluminal propagation, and without time delay is proposed and developed in $arXiv:1005.3999V1$ in 2010.
In the paper, we present the GLLH cloaks imager in Figure 1 and the GILD cloak phenomena imager in Figure 2; both match very well.
\begin{figure}[h]
\centering
\includegraphics[width=0.9\linewidth,draft=false]{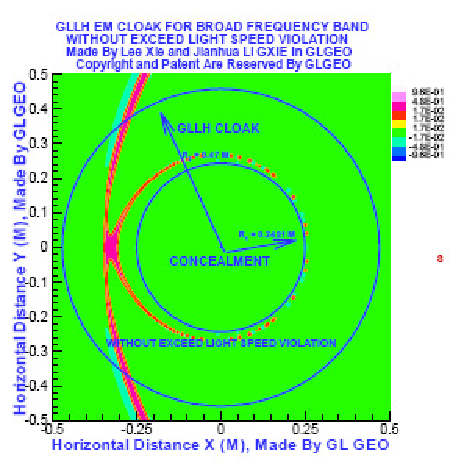}
\caption{ (color online) 
Electric wave propagation at time step $119dt$, the two peaks of
the crescent like wave front intersects at a branching point, red S in right is source location}\label{Fig1}
\end{figure}
\begin{figure}[b]
\centering
\includegraphics[width=0.86\linewidth,draft=false]{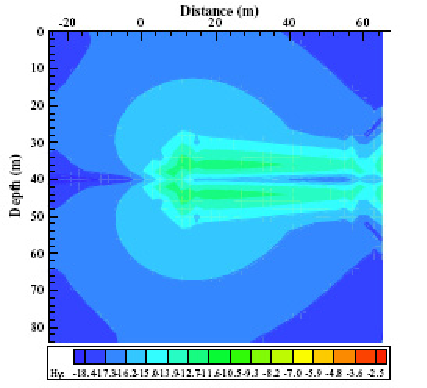}
\caption{ (color online) 
A novel cloak surround space shuttle model to prevent detection has been discovered in 
GILD EM inversion in 2001, which was published in SEG expand abstract in 2002[7]
.}\label{fig2}
\end{figure}

The GLLH invisible cloak is not from coordinate transformation. Find an invisible cloak is the non scattering problem. Using Global and Local (GL) electromagnetic non-scattering modeling and inversion with a distinctive class of materials is one method in our paper in $arXiv:1005.3999V1$ in 2010. Another method is some 0 to R1 radial coordinate transformations with superluminal and infinite phase velocity. From the GLLH invisible cloak, we discovered positive space and negative space and invisible science[8]. Using a new negative infinity to 0 quasi topological transformation, we discovered new isotropic electromagnetic invisible GLHUA sphere[9]; anisotropic electromagnetic invisible GLHUA cloak without superluminal and time delay[10-13].
In the GLLH cloak[1], the wave front is curved as a crescent like and without superluminal and time delay. Open question: Can we construct a 3D Kakeya set (see Guth, Hong Wang and Joshua Zahl [25][26]), where line segments of length  represent the vector field of EM wave ray-tracing propagation through GLLH[1], GLHUA cloaks[9], and the GLHUA sphere[10]? 

From the GLLH invisible cloak, we discovered super science[8], positive space and negative space and invisible science; isotropic electromagnetic invisible GLHUA sphere; anisotropic electromagnetic invisible GLHUA cloak[8][9][10]. What is the positive space and negative space? Just as before, without imaginary number, i=?(-1), we cannot solve the equation  in the real number domain. Let's ask: In polar coordinates on a two-dimensional plane, the , , is a point on the plane, where is  for the same value? This question can intrigue students from elementary middle school to university, and has also attracted the attention of professors and scientists at all levels. In this article, we introduce the concepts of positive space and negative space. In the three-dimensional spherical coordinate system, the set of points with positive radial coordinates,,, ,is called three-dimensional positive space. This is the three-dimensional space we live in. In the three-dimensional spherical coordinate system, a point with negative radial coordinates, , ,for same, is defined as three-dimensional negative space, as described in our paper [8][9][10][11][12][13]. The origin of the positive space can be anywhere. The point in the positive space, , and the point in the negative space, , are corresponding points with the same longitude and latitude. The origin of the positive space, +0, and the origin of the negative space, -0, are corresponding points, but different points. The three-dimensional negative space is the corresponding space of the three-dimensional positive space. When we are awake, we act in all kinds of strange ways in the positive space; when we are asleep, our sweet dreams are in the negative space. The GLLH invisible cloak is not from coordinate transformation. A quasi  topological coordinate transform maps the outside negative sphere, , in negative space to annular,  , in positive space to develop  an adavance AGLLH invisible cloak is proposed in the section 7.
The structure of this paper is as follows: The introduction is in Section 1. In Section 2, we propose a new GLLH EM cloak without exceeding the speed of light. The GL EM cloak inversion is proposed in Section 3. In Section 4, we present novel wavefront branching and EM wave propagation in the GLLH cloak without exceeding the speed of light. The distinct invisibility properties of the GLLH EM cloak are discussed in Section 5. In Section 6, we propose a GL sphere harmonic modeling and inversion. Section 7 describes an advanced GLLH Invisibility AGLLH Cloak . The discussion and conclusion is in Section 8 .

\section {GLLH EM CLOAK WITHOUT EXCEEDING LIGHT SPEED}
In this section, we propose a new GLLH EM invisible cloak without exceeding the light speed violation
\subsection{GLLH EM INVISIBLE CLOAK}
In the Maxwell equation
\begin{equation}
\begin{array}{l}
 \nabla  \times \vec E =  - \bar \mu \frac{\partial }{{\partial t}}\vec H + \vec M \\ 
 \nabla  \times \vec H = \bar \varepsilon \frac{\partial }{{\partial t}}\vec E + \vec J. \\ 
 \end{array}
\end{equation}

\[
\nabla \times \vec{E} = \frac{1}{r^2 \sin \theta} \begin{bmatrix} \vec{r} & r \vec{\theta} & r \sin \theta \vec{\phi} \\ \dfrac{\partial}{\partial r} & \dfrac{\partial}{\partial \theta} & \dfrac{\partial}{\partial \phi} \\ E_r & r E_{\theta} & r \sin \theta E_{\phi} \end{bmatrix}\]

Where $\vec E $ is  eletric wave vector field,$\vec H $ is magbetic wave vector field 

\begin{equation}
\bar \varepsilon  = diag\left[ {\begin{array}{*{20}c}
   {\varepsilon _r } & {\varepsilon _\theta  } & {\varepsilon _\phi  }  \\
\end{array}} \right]\varepsilon _0 ,R_i  \le r \le R_o,
\end{equation}
\begin{equation}
\bar \mu  = diag\left[ {\begin{array}{*{20}c}
   {\mu _r } & {\mu _\theta  } & {\mu _\phi  }  \\
\end{array}} \right]\mu _0 ,R_i  \le r \le R_o.
\end{equation}
We propose the GLLH EM invisible cloak material parameters in a 3D spherical coordinate system. The cloak material parameters are radial dependent as follows: 

\begin{equation}
\begin{array}{l}
 \varepsilon _r  = \mu _r  = \frac{1}{{1 + 4R_o  + 2R_o^2 }}\left( {\frac{{2R_o^2 \left( {R_o  - R_i } \right)}}{{r^2 }}} \right. \\ 
  + \frac{{2 + 6R_o }}{{\log \left( {e^{1/R_o } \left( {R_0  - R_i } \right)/\left( {r - R_i } \right)} \right)}}\frac{{\left( {r - R_i } \right)}}{{R_o r^2 }} \\ 
 \left. { - \frac{{1 + 2R_o }}{{\log ^2 \left( {e^{1/R_o } \left( {R_o  - R_i } \right)/\left( {r - R_i } \right)} \right)}}\frac{{\left( {r - R_1 } \right)}}{{R_o^2 r^2 }}} \right), \\ 
 \end{array}
\end{equation}
\begin{equation}
\begin{array}{l}
 \varepsilon _\theta   = \varepsilon _\phi   = \mu _\theta   = \mu _\phi   \\ 
  = \frac{R_o}{{2\log \left( {e^{1/R_o } \left( {R_0  - R_i } \right)/\left( {r - R_i } \right)} \right)\left( {r - R_i } \right)}} \\ 
  + \frac{1}{{2R_o\log ^3 \left( {e^{1/R_o } \left( {R_0  - R_i } \right)/\left( {r - R_i } \right)} \right)\left( {r - R_i } \right)}} \\ 
 \end{array}
\end {equation}

where the $R_i  \le r \le R_o$ is the spherical annular cloak 
domain,  $R_i$ is inner radius of the annular cloak,
$R_o$ is outer radius of the annular cloak, 
${\bar \varepsilon }$ is the dielectric parameter matrix,  ${\bar \mu }$ is the permeability  parameter matrix, 
$\varepsilon _0 $ is the basic dielectric parameter, $\mu _0 $ is the basic magnetic permeability parameter, ${\varepsilon _r }$ is the relative radial dielectric permittivity, ${\varepsilon _\theta  }$ and ${\varepsilon _\phi  }$  are the relative angular dielectric permittivity,
${\mu _r }$ is relative radial permeability, ${\mu _\theta  }$ and ${\mu _\phi  }$ are relative angular permeability. The source free in the GLLH invisible cloak.

\subsection{INVISIBLE FUNCTIONS OF GLLH EM CLOAK }
\subsubsection{function I}
When the source $r_s$ and observer $r$ are  located outside of the cloak, we verify the invisible function I of the GLLH EM cloak, $n^2(R_o) = \mu_{\theta} * \varepsilon_r(R_o) = 1$ , Non scattering from the GLLH cloak to disturb the exterior EM field outside GLLH cloak with materials in (4)-(5) as follows: 

\hfill\break

$Exterior\ EM\ field\ must\ not\ be \ interfered \ by\  scattering $ 
   \ $ from \ the\  cloak:$

\begin{equation}
\left[ {\begin{array}{*{20}c}
   {E(r,r_s ,t)}  \\
   {H(r,r_s ,t)}  \\
\end{array}} \right] = \left[ {\begin{array}{*{20}c}
   {E_b (r,r_s ,t)}  \\
   {H_b (r,r_s ,t)}  \\
\end{array}} \right],
\end{equation}
\subsubsection{function II}
When the source $r_s$ is located outside of the cloak and observer $r$ is inside of the cloak, we verify the invisible function II of the GLLH EM cloak with materials in (4)-(5) as follows: 
\[{Exterior\ EM\ field\ does\ not\ penetrate\ into\ concealment:}\]    
\begin{equation}
\left[ {\begin{array}{*{20}c}
   {E(r,r_s ,t)}  \\
   {H(r,r_s ,t)}  \\
\end{array}} \right] = 0,
\end{equation}
By (4) and (5),$ n(r) = \mu_{\theta} \varepsilon_r(r) \geq 1$, the GLLH EM cloak completely overcomes the  problems on $exceeding \ light \ speed$ and   $infinite \ speed$
two physical violations. 

\section {GL EM  CLOAK INVERSION}

We proposal a GL EM inversion for cloak in this section. It is based on the 3D EM integral equation and GL EM modeling in [3] and GL Metro Carlo inversion in [5].

\subsection{3D EM INTEGRAL EQUATION}
  
We propose a GL EM inversion for GLLH cloak in this section. It is based on the 3D EM integral equation and GL EM modeling in [ 3] and the GL Metro Carlo inversion in [ 5 ].
A 3D electromagnetic integral equation is proposed in this section as follow:
\begin{equation}
\begin{array}{l}
 \left[ {\begin{array}{*{20}c}
   {E(r,r_s ,t)}  \\
   {H(r,r_s ,t)}  \\
\end{array}} \right] = \left[ {\begin{array}{*{20}c}
   {E_b (r,r_s ,t)}  \\
   {H_b (r,r_s ,t)}  \\
\end{array}} \right] +  \\ 
  + \int\limits_{\Omega _{cloak} } {G_{E,H,b}^{J,M} } (r',r,t) * _t \left[ {\begin{array}{*{20}c}
   {\delta D_{11} } & {}  \\
   {} & {\delta D_{22} }  \\
\end{array}} \right]\left[ {\begin{array}{*{20}c}
   {E(r',r_s ,t)}  \\
   {H(r',r_s ,t)}  \\
\end{array}} \right], \\ 
 \end{array}
\end{equation}

\begin{equation}
\begin{array}{l}
 \left[ {\begin{array}{*{20}c}
   {E(r,r_s ,t)}  \\
   {H(r,r_s ,t)}  \\
\end{array}} \right] = \left[ {\begin{array}{*{20}c}
   {E_b (r,r_s ,t)}  \\
   {H_b (r,r_s ,t)}  \\
\end{array}} \right] +  \\ 
  + \int\limits_{\Omega _{cloak} } {G_{E,H}^{J,M} } (r',r,t) * _t \left[ {\begin{array}{*{20}c}
   {\delta D_{11} } & {}  \\
   {} & {\delta D_{22} }  \\
\end{array}} \right]\left[ {\begin{array}{*{20}c}
   {E_b (r',r_s ,t)}  \\
   {H_b (r',r_s ,t)}  \\
\end{array}} \right], \\ 
 \end{array}
\end{equation}

Where  $ E(r,r_s ,t)$ is the electric intensity field,  $ H(r,r_s ,t)$   the is magnetic intensity field, $r$,  is observe space location variable,  $r_s$  is source  space location variable, $t$ is time, $ E_b(r,r_s ,t)$  is the background electric intensity field, $ H_b(r,r_s ,t)$  is the background  magnetic intensity field, $ G_{E,H,b}^{J,M} (r?r,t) $ is $ 6 \times 6  $ EM $Green's$ tensor matrix, $diagonal\left[ {\delta D_{11} ,\delta D_{22} } \right] $  is is $ 6 \times 6$  variation EM material matrix as follows:

\begin{equation}
\begin{array}{l}
 \delta D_{11}  = (\bar \varepsilon  - \varepsilon _0 I)\frac{\partial }{{\partial t}},{\rm  }\delta D_{22}  = (\bar \mu  - \mu _0 I)\frac{\partial }{{\partial t}} \\ 
 \bar \varepsilon  = diag\left[ {\begin{array}{*{20}c}
   {\varepsilon _r } & {\varepsilon _\theta  } & {\varepsilon _\phi  }  \\
\end{array}} \right]\varepsilon _0 ,{\rm  } \\ 
 \bar \mu  = diag\left[ {\begin{array}{*{20}c}
   {\mu _r } & {\mu _\theta  } & {\mu _\phi  }  \\
\end{array}} \right]\mu _0 , \\ 
 \end{array}
\end{equation}
Where ${\bar \varepsilon }$ is the dielectric parameter matrix,  ${\bar \mu }$ is the permeability  parameter matrix, the dielectric and permeability can be isotropic or anisotropic materials, $ \varepsilon _0 $ is the basic dielectric parameter, $\mu _0 $ is
the basic magnetic permeability parameter, ${\varepsilon _r }$ is relative dielectric 
in $r$ direction, ${\varepsilon _\theta  }$ is relative dielectric 
in $\theta  $ direction, ${\varepsilon _\phi  }$ is relative dielectric in $\phi $ direction,
${\mu _r }$ is relative permeability 
in $r$ direction, ${\mu _\theta  }$ is relative permeability 
in $\theta  $ direction, ${\mu _\phi  }$ is relative permeability in $\phi $ direction, for the  cloak material, the ${\bar \varepsilon }$ and ${\bar \mu }$ are proposed
in formulas  $(3) - (5)$ in this paper.

\subsection{FUNCTIONS IN GL EM CLOAK INVERSION }

We propose a GLLH EM cloak inversion for invisible cloak in this section. We request two invisible functions and without
exceed light speed violation for the GL EM cloak:  (I) the exterior EM field should not be scattering interfered by the cloak ;  (II) the exterior EM field can not penetrate into the concealment of the cloak; and (III) without exceed light speed violation. 

\subsection {GL EM CLOAK EXTERIOR INVERSION}
When the source $r_s$ and observer $r$ are  located outside of the cloak, we present the cloak invisible function I as the following inversion.
\hfill\break

$Exterior\ EM\ field\ must\ not\ be \ interfered \ by \ scattering $ 
   \ $ from \ the\  cloak.$ 
\\\

According to the invisible function I (6), and using the EM integral equation (8), we have

\begin{equation}
\int\limits_{\Omega _{cloak} } {G_{E,H,b}^{J,M} } (r',r,t) * _t \left[ {\begin{array}{*{20}c}
   {\delta D_{11} } & {}  \\
   {} & {\delta D_{22} }  \\
\end{array}} \right]\left[ {\begin{array}{*{20}c}
   {E(r',r_s ,t)}  \\
   {H(r',r_s ,t)}  \\
\end{array}} \right] = 0,
\end{equation}

Using the EM integral equation (9), we have

\begin{equation}
\int\limits_{\Omega _{cloak} } {G_{E,H}^{J,M} } (r',r,t) * _t \left[ {\begin{array}{*{20}c}
   {\delta D_{11} } & {}  \\
   {} & {\delta D_{22} }  \\
\end{array}} \right]\left[ {\begin{array}{*{20}c}
   {E_b(r',r_s ,t)}  \\
   {H_b(r',r_s ,t)}  \\
\end{array}} \right] = 0,
\end{equation}
with a distinctive class of materials $ a_{\alpha \beta}\log ^\alpha (b_{\alpha \beta}/h) h^\beta $ [3][4][5]. 

\subsection {GL EM CLOAK INNER INVERSION}

When the source $r_s$ is located outside of the cloak and observer $r$ is inside of the cloak , we present the cloak invisible function II in the following inversion. 
\[{Exterior\ EM\ field\ does\ not\ penetrate\ into\ concealment.}\]    
\ \ \
According to the invisible function II (7) and using the EM integral equation (8), we have
\begin{equation}
\begin{array}{l}
 \left[ {\begin{array}{*{20}c}
   {E_b (r,r_s ,t)}  \\
   {H_b (r,r_s ,t)}  \\
\end{array}} \right] \\ 
  + \int\limits_{\Omega _{cloak} } {G_{E,H,b}^{J,M} } (r',r,t) * _t \left[ {\begin{array}{*{20}c}
   {\delta D_{11} } & {}  \\
   {} & {\delta D_{22} }  \\
\end{array}} \right]\left[ {\begin{array}{*{20}c}
   {E(r',r_s ,t)}  \\
   {H(r',r_s ,t)}  \\
\end{array}} \right] = 0, \\ 
 \end{array}
\end{equation}
By using the EM integral equation (9), we have

\begin{equation}
\begin{array}{l}
 \left[ {\begin{array}{*{20}c}
   {E_b (r,r_s ,t)}  \\
   {H_b (r,r_s ,t)}  \\
\end{array}} \right] +  \\ 
 \int\limits_{\Omega _{cloak} } {G_{E,H}^{J,M} } (r',r,t) * _t \left[ {\begin{array}{*{20}c}
   {\delta D_{11}\ 0}  \\
   {0 \ \delta D_{22} }  \\
\end{array}} \right]\left[ {\begin{array}{*{20}c}
   {E_b(r',r_s ,t)}  \\
   {H_b(r',r_s ,t)}  \\
\end{array}} \right] = 0, \\ 
 \end{array} 
\end{equation}
with a distinctive class of materials $ a_{\alpha \beta}\log ^\alpha (b_{\alpha \beta}/h) h^\beta $ [3][4][5]. the $(8)-(14)$ above can be simplified by the spherical harmonic analysis in Section 6.

\subsection {CONSTRAINT IN GLLH EM CLOAK INVERSION}

We present the without exceeding light speed violation as the constraint of the GLLH EM cloak inversion,
for the radial dependent ${\bar \varepsilon (r)} $ and ${\bar \mu(r)}$,   
\begin {equation}
\begin{array}{l}
 \varepsilon _r (r) \mu _\theta (r)  \ge 1, \\ 
 \varepsilon _\theta (r) \mu _r (r)  \ge 1, \\ 
 \varepsilon _\theta (r)  = \varepsilon _\phi (r)  = \mu _\theta (r)   = \mu _\phi  (r).  \\ 
 \end{array}
\end{equation}

\subsection {GL EM CLOAK INVERSION}

The EM integral equation  (11) and (12) for $r_s  > R_o$, and $r > R_o 
$, the EM integral equation (13) and (14) for $r_s  > R_o$, and $r < R_i$, and the "No exceed light speed" radial dependent constraint (15) are coupled to construct the GL EM cloak inversion for the EM invisible cloak. We use the GL EM Metro Carlo inversion method in [3] to
solve the GLLH EM cloak inversion, we find the radial dependent GLLH EM cloak
material $ a_{\alpha \beta}\log ^\alpha (b_{\alpha \beta}/h) h^\beta $[3][4][5] to satisfy the GL EM cloak exterior inversion (11) and(12); the GL EM cloak inner inversion (13) and (14); and  constraint (15) without exceeding light violation.

\begin{figure}[b]
\centering
\includegraphics[width=0.86\linewidth,draft=false]{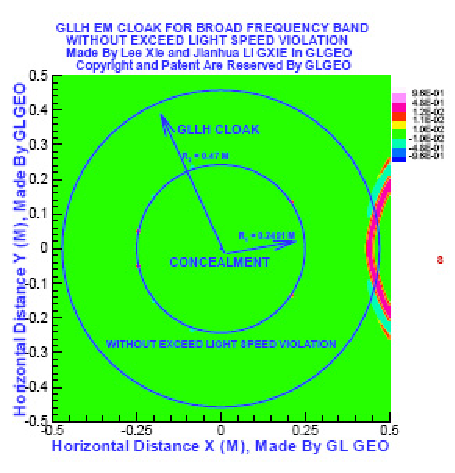}
\caption{ (color online) 
At time step $40dt$, 
front of $ {\it   Electric\  wave} $, $E_{x}$ inside of the  GLLH EM cloak $R_i \le r \le R_o$
propagates no faster than light speed. The red S in right is
source.}\label{fig3}
\end{figure}
\begin{figure}[b]
\centering
\includegraphics[width=0.86\linewidth,draft=false]{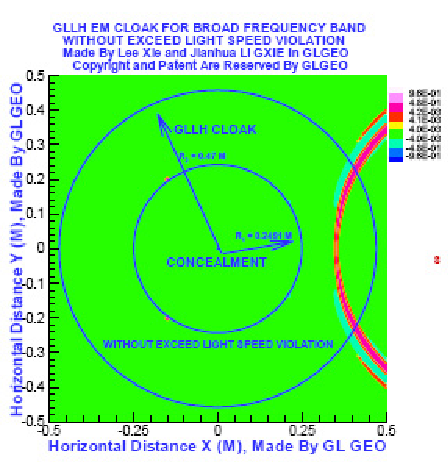}
\caption{ (color online)
At time step $50dt$, 
front of $ {\it   Electric\  wave} $, $E_{x}$ inside of the  GLLH EM cloak $R_i \le r \le R_o$
propagates no faster than light speed. the red S in right is
source.  }\label{fig4}
\end{figure}
\begin{figure}[b]
\centering
\includegraphics[width=0.86\linewidth,draft=false]{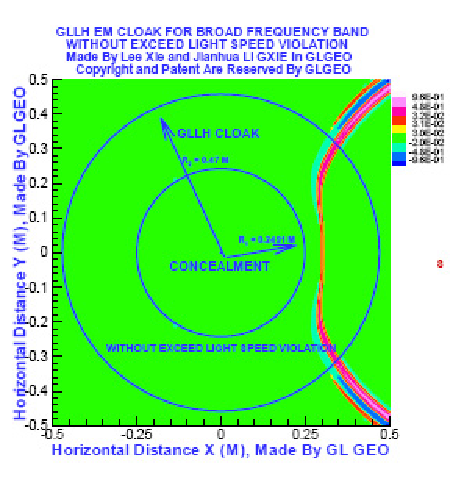}
\caption{ (color online) 
At time step $60dt$, 
front of $ {\it   Electric\  wave} $, $E_{x}$ inside of the  GLLH EM cloak $R_i \le r \le R_o$
propagates slower than light speed. the red S in right is
source.}\label{fig5}
\end{figure}
\begin{figure}[h]
\centering
\includegraphics[width=0.86\linewidth,draft=false]{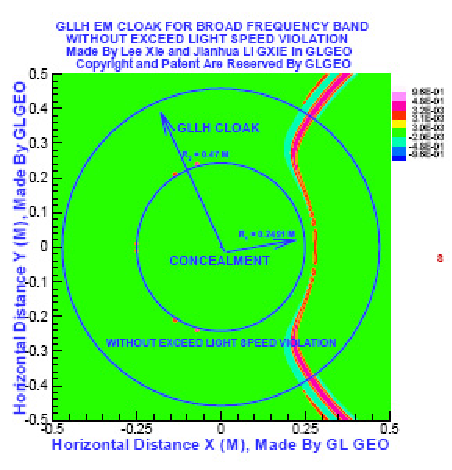}
\caption{ (color online) 
At time step $70dt$, 
front of $ {\it   Electric\  wave} $, $E_{x}$ inside of the  GLLH EM cloak $R_i \le r \le R_o$
propagates slower than light speed, dashed
wavefront bowing to the right indicates that the
incident wavefront is absorbed.}\label{fig6}
\end{figure}
\begin{figure}[h]
\centering
\includegraphics[width=0.85\linewidth,draft=false]{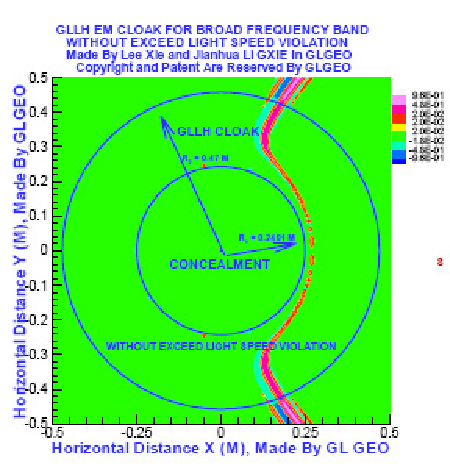}
\caption{ (color online) 
At time step $80dt$, 
front of $ {\it   Electric\  wave} $, $E_{x}$ inside of the  GLLH EM cloak $R_i \le r \le R_o$
propagates slower than light speed. The dashed
points right behind denote discontinuous
absorbing propagation. }\label{fig7}
\end{figure}
\begin{figure}[h]
\centering
\includegraphics[width=0.85\linewidth,draft=false]{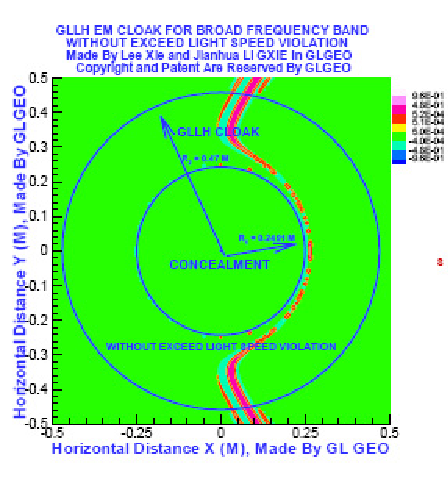}
\caption{ (color online)
At time step $90dt$, 
front of $ {\it   Electric\  wave} $, $E_{x}$ inside of the  GLLH EM cloak $R_i \le r \le R_o$
propagates slower than light speed. The dashed
points right behind denote discontinuous
absorbing propagation.}\label{fig8}
\end{figure}
\begin{figure}[h]
\centering
\includegraphics[width=0.85\linewidth,draft=false]{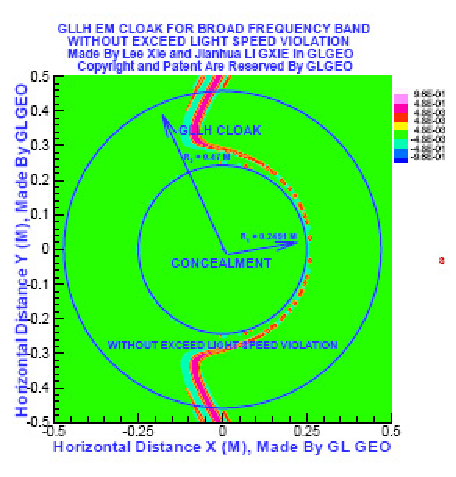}
\caption{ (color online) 
At time step $100dt$, 
front of $ {\it   Electric\  wave} $, $E_{x}$ inside of the  GLLH EM cloak $R_i \le r \le R_o$
propagates slower than light speed. The dashed
points right behind denote discontinuous
absorbing propagation, left wave front is created
wave discontinuously. }\label{fig9}
\end{figure}
\begin{figure}[h]
\centering
\includegraphics[width=0.85\linewidth,draft=false]{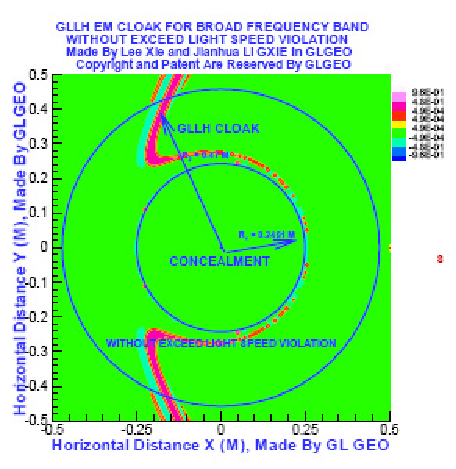}
\caption{ (color online) 
At time step $110dt$, 
front of $ {\it   Electric\  wave} $, $E_{x}$ inside of the  GLLH EM cloak $R_i \le r \le R_o$
propagates slower than light speed. The dashed
points right behind denote discontinuous
absorbing propagation, left wave front is created
wave, discontinuously.}\label{fig10}
\end{figure}
\begin{figure}[h]
\centering
\includegraphics[width=0.85\linewidth,draft=false]{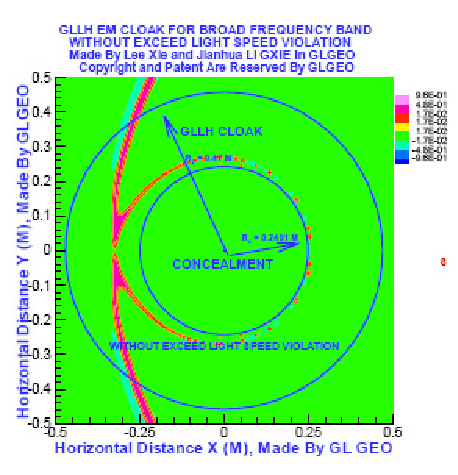}
\caption{ (color online) 
At time step $118dt$, 
front of $ {\it   Electric\  wave} $, $E_{x}$ inside of the  GLWF cloak $R_i \le r \le R_o$
propagates slower than light speed. Upper
downward break mouth and Lower upward break
mouth will close, the created wave likes a
crescent. The dashed points right behind denote
discontinuous absorbing propagation, left wave
front is created wave, discontinuously. }\label{fig11}
\end{figure}
\begin{figure}[h]
\centering
\includegraphics[width=0.85\linewidth,draft=false]{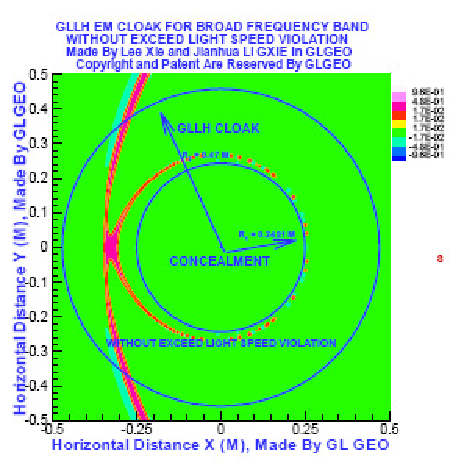}
\caption{ (color online) 
At time step $119dt$, 
front of $ {\it   Electric\  wave} $, $E_{x}$ inside of the  GLLH EM cloak $R_i \le r \le R_o$
propagates slower than light speed. The curved front intersects
at branching point in left which depends on the source. Upper downward break mouth and Lower
upward break mouth closed, the created wave
likes a crescent . The dashed points right behind
denote discontinuous absorbing propagation, left
wave front is created wave, discontinuously. The red S in right is source.  }\label{fig12}
\end{figure}
\begin{figure}[h]
\centering
\includegraphics[width=0.85\linewidth,draft=false]{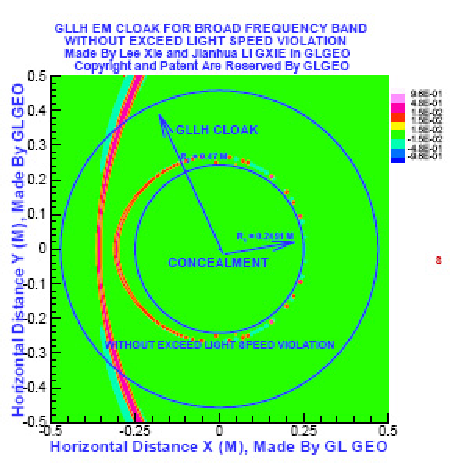}
\caption{ (color online) 
At time step $120dt$, 
front of $ {\it   Electric\  wave} $, $E_{x}$ inside of the  GLLH EM cloak $R_i \le r \le R_o$
propagates slower than light speed. The curved front is split to 
two branching fronts. The
crescent front is split into two branching fronts .
The created left front propagates outward. The
dashed points on the right rear and left to right
dashed front form a discontinuous absorbing circle
propagation. }\label{fig13}
\end{figure}
\begin{figure}[h]
\centering
\includegraphics[width=0.85\linewidth,draft=false]{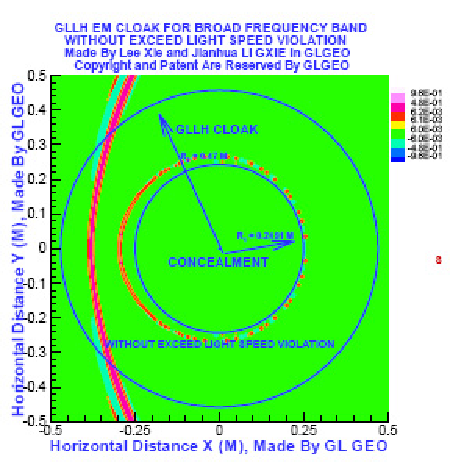}
\caption{ (color online) 
At time step $122dt$, 
front of $ {\it   Electric\  wave} $, $E_{x}$ inside of the  
crescent front is split into two branching fronts .
The created left front propagates outward. The
dashed points on the right rear and left to right
dashed front form a absorbing circle of attractive
propagation. }\label{fig14}
\end{figure}
\begin{figure}[h]
\centering
\includegraphics[width=0.85\linewidth,draft=false]{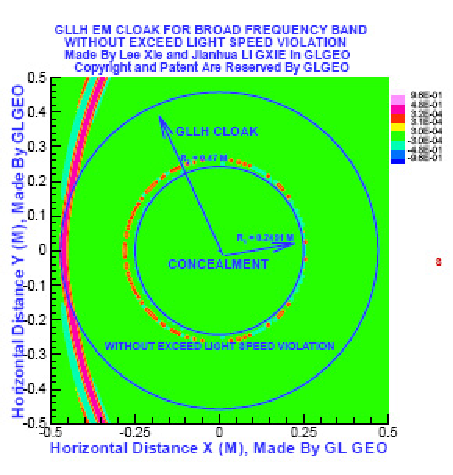}
\caption{ (color online)
At time step $130dt$, 
front of $ {\it   Electric\  wave} $, $E_{x}$ inside of the  GLLH EM cloak $R_i \le r \le R_o$ propagates slower than light speed. The left
front propagates outward arrive to left boundary
without time delay, The dashed points on the right
rear and left to right front form an absorbing circle
of attractive propagation. }\label{fig15}
\end{figure}
\begin{figure}[h]
\centering
\includegraphics[width=0.85\linewidth,draft=false]{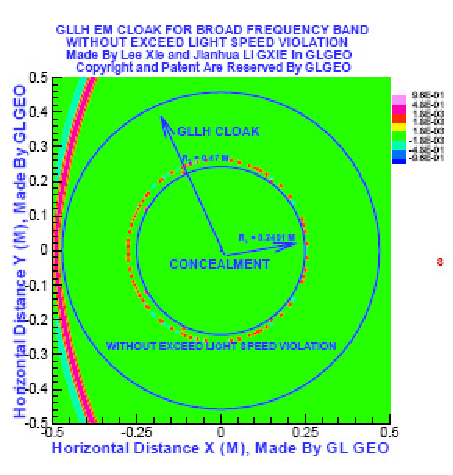}
\caption{ (color online) 
At time step $133dt$, 
front of $ {\it   Electric\  wave} $, $E_{x}$ inside of the  GLLH EM cloak $R_i \le r \le R_o$
propagates slower than light speed. The left
front propagates outward arrive left boundary
with time delay, The dashed points on the right
rear and left to right front form an absorbing
circle of attractive to inner circle.}\label{fig16}
\end{figure}
\begin{figure}[h]
\centering
\includegraphics[width=0.85\linewidth,draft=false]{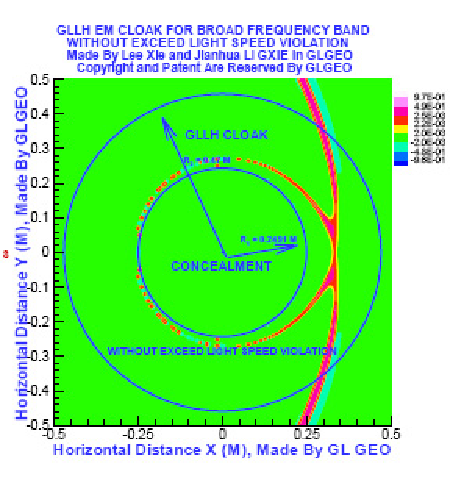}
\caption{ (color online) 
At time step $119dt$, 
front of $ {\it   Electric\  wave} $, $E_{x}$ inside of the  GLLH EM cloak $R_i \le r \le R_o$
propagates slower than light speed. Upper downward break mouth
and lower upward break mouth closed, the
created wave like a crescent . The dashed points
left denote discontinuous absorbing propagation,
discontinuous absorbing propagation, right wave
front is created wave, discontinuously. The red S in left is
source.  }\label{fig17}
\end{figure}
\begin{figure}[h]
\centering
\includegraphics[width=0.86\linewidth,draft=false]{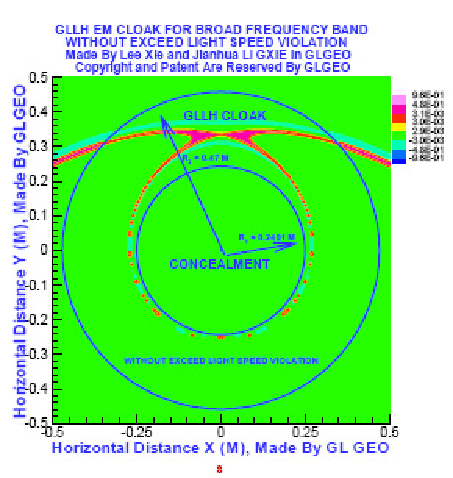}
\caption{ (color online) 
At time step $119dt$, 
front of $ {\it   Electric\  wave} $, $E_{x}$ inside of the  GLLH EM cloak $R_i \le r \le R_o$
propagates slower than light speed. Left rightward break mouth
and right leftward break mouth closed, the
created wave like a crescent. The dashed points
botton denote discontinuous absorbing
propagation, up wave front is created wave,
discontinuously}\label{fig18}
\end{figure}

\begin{figure}[h]
\centering
\includegraphics[width=0.86\linewidth,draft=false]{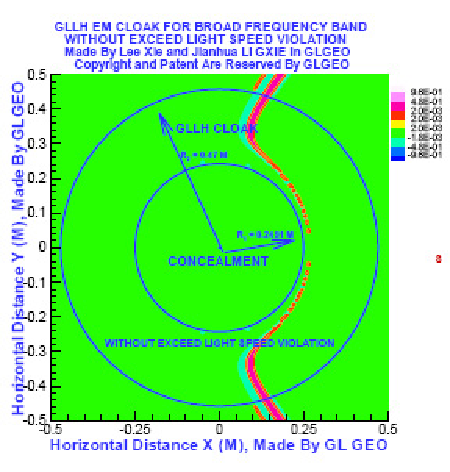}
\caption{ (color online) 
At time step $84dt$, 
front of $ {\it   Electric\  wave} $, $E_{x}$ inside of the  GLLH EM cloak $R_i \le r \le R_o$
propagates slower than light speed. The dashed points right behind denote discontinuous absorbing propagation. This indicates  the wave propagation in GLLH cloak without superluminal problem. The red source S located on the right. The red S in right is
source.  }\label{fig19}
\end{figure}
\begin{figure}[h]
\centering
\includegraphics[width=0.86\linewidth,draft=false]{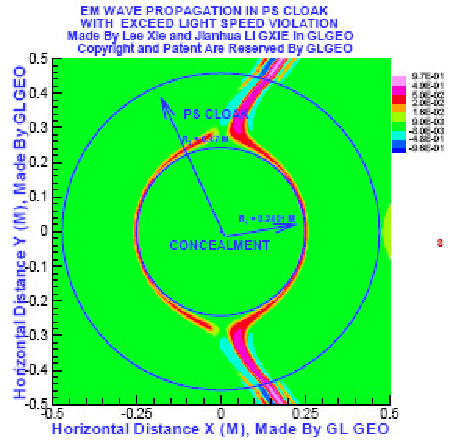}
\caption{ (color online) 
At time step $84dt$, 
front of $ {\it   Electric\  wave} $, $E_{x}$ inside of the  Pendry  cloak $R_i \le r \le R_o$.  When wave arrives at r=Ri,  the front forms closed circle around the inner boundary . This indicates, on inner circle r=Ri, the wave propagation in Pendry cloak has infinite speed. The red source S located on the right.
 }\label{fig20}
\end{figure}
\begin{figure}[h]
\centering
\includegraphics[width=0.85\linewidth,draft=false]{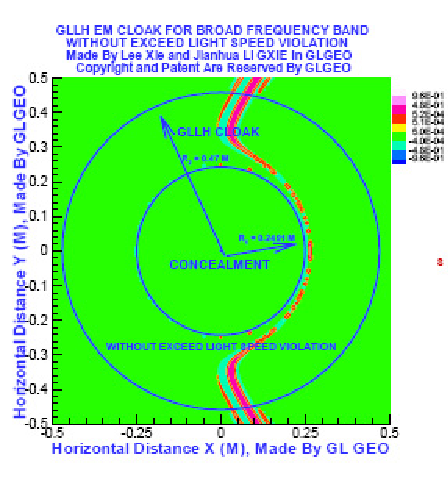}
\caption{ (color online)
At time step $90dt$, 
front of $ {\it   Electric\  wave} $, $E_{x}$ inside of the  GLLH EM cloak $R_i \le r \le R_0$
propagates slower than light speed. The dashed points right behind denote discontinuous absorbing propagation. This indicates the wave propagation in GLLH cloak without superluminal problem. The red source S is located on the right.}\label{fig21}
\end{figure}
\begin{figure}[h]
\centering
\includegraphics[width=0.86\linewidth,draft=false]{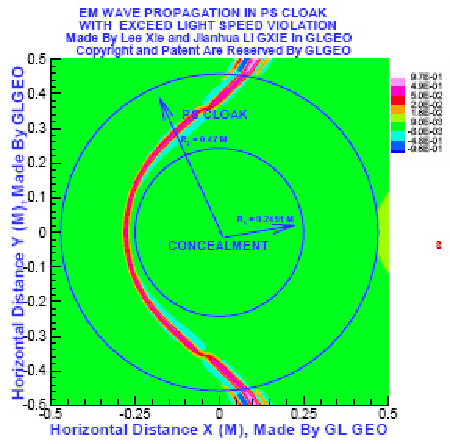}
\caption{ (color online) 
At time step $90dt$, 
front of $ {\it   Electric\  wave} $, $E_{x}$ inside of the  Pendry cloak $R_i \le r \le R_o$
propagation is much more exceed light speed. The red S in righ  is source. The electromagnetic wave propagation in Pendry cloak has superluminal propagation.}\label{fig22}
\end{figure}
\section{NOVEL EM PROPAGATION IN GLLH INVISIBLE CLOAK WITHOUT EXCEEDING LIGHT SPEED }

In this section, we present a novel EM propagation in the EM GLLH invisible cloak without exceeding light speed. The EM wave propagation pattern is completely new in [1], arXiv:1005.3999, and has never been shown in other authors' papers published before arXiv:1005.3999 in May 2010. GLLH invisible cloak is not from the coordinate transformation.

\subsection{ THE SIMULATION MODEL OF THE EM GLLH CLOAK }
The simulation model: the 3D domain is $ [-0.5m,0.5 m] \times [-0.5m,0.5 m]  \times [-0.5m,0.5m]$, 
the mesh number is $201 \times 201 \times 201$, the mesh size is 0.005m. 
The electric current point source is defined as
\begin{equation}
\delta (r - r_s )\delta (t)\vec e,
\end{equation}
where the $r_s > R_o $ denotes the location of the point source,
the unit vector $\vec e$ is the polarization direction, 
the time step $dt = 0.3333 \times 10^{ - 10}$ 
second, the frequency band is from 0.05 GHz to 15 GHz, the  largest frequency $f=15 GHz$,  
the shortest wave length is $0.02m$. The GLLH EM cloak is consist of the spherical annular $R_i  \le r \le R_o$ with center in original point
and is fill in by the GLLH EM cloak materials , where 
$ R_i=0.2491\ meter$, $R_o=0.47\ meter$.  The cloak is divided into 
$90 \times 180 \times 90$ cells. 
The spherical coordinate is used in the sphere 
$r \le R_o$, the Cartesian rectangular coordinate is used in outside the sphere $r\ >\ R_o$ to mesh the domain. 

\subsection {NOVEL ELECTRIC WAVE PROPAGATION}

The electric intensity wave is excited by the point source S, denoted by a red S in the figures in this paper. In Figure 1 and Figures $3 - 16$, the source ( red S ) is located in free space, on the right side outside of the entire GLLH EM cloak, at ( 0.83 m , 0. 0, 0. 0 ).  In Figure 17, the point source ( red S ) is located on the left outer side of the point source marked by the red S is located on the left outer side of the cloak, at (- 0.83 m, 0. 0, 0. 0 ). In Figure 18, it is located on the bottom outer side the cloak, at ( 0. 0, -0.83m , 0. 0 ).  In Figure 3, at time step 40 dt, the electric wave, Ex, inside the GLLH EM cloak, 
$R_i \le r \le R_o$
,
propagates no faster than the speed of light. Figure 4 shows that at time step 50 dt, the electric wave, Ex, inside the GLLH EM cloak , 
$R_i \le r \le R_o$
,
 propagates no faster than the speed of light. In Figure 5, at time step 60 dt , the electric wave, Ex, inside the GLLH EM cloak R1 ? r ? R2 propagates slower than the speed of light.  In Figure 6, At time step 70dt, the front of electric wave, Ex inside the GLLH EM cloak, 
$R_i \le r \le R_o$, 
propagates slower than speed of light. dashed wavefront bowing to the right indicates that the incident wavefront is absorbed. In Figure 7, At time step 80dt, the front of electric wave, Ex inside the GLLH EM cloak ,
$R_i \le r \le R_o$,
propagates slower than speed of light. The dashed points right behind denote discontinuous absorbing propagation. Figure 8 shows that  at time step 90dt, the front of electric wave, Ex inside the GLLH EM cloak 
$R_i \le r \le R_o$

propagates slower than speed of light. The dashed points right behind denote discontinuous absorbing  propagation. In Figure 9, at time step 100dt, the front of electric wave, Ex  inside the GLLH EM cloak 
$R_i \le r \le R_o$

propagates slower than speed of light. The dashed points right behind denote discontinuous absorbing  propagation, left wave front  is created wave discontinuously. In Figure 10, at time step 110dt, the front of electric wave, Ex  inside the GLLH EM cloak 
$R_i \le r \le R_o$ 
propagates slower than speed of light. The dashed points right behind denote discontinuous absorbing propagation, left wave front  is created wave, discontinuously. Figure 11 shows that at time step 118dt, the front of electric wave, Ex inside the GLLH EM cloak
 $R_i \le r \le R_o$
propagates slower than speed of light. Upper downward break mouth and Lower upward break mouth will close, the created wave likes a crescent. The dashed points right behind denote discontinuous absorbing propagation, left wave front is created wave, discontinuously. In Figure 1 and 12, at time step 119dt, the front of electric wave, Ex inside the GLLH EM cloak, 
$R_i \le r \le R_o$,
propagates slower than speed of light. Upper downward break mouth and Lower upward break mouth closed, the created wave likes  a crescent . The dashed points right behind denote discontinuous absorbing propagation, left wave front  is created wave, discontinuously. In Figure 13, At time step 120dt, the front of electric wave, Ex inside the GLLH EM cloak, 
$R_i \le r \le R_o$, 
propagates slower than speed of light. The crescent front is split into two branching fronts  . The created left front propagates outward. The dashed points on the right rear and left to right dashed front form a discontinuous absorbing circle propagation. In Figure 14, at time step 122dt, the front of electric wave, Ex inside the GLLH EM cloak, 
$R_i \le r \le R_o$,
propagates slower than speed of light. The crescent front is split into two branching fronts . The created left front propagates outward. The dashed points on the right rear and left to right dashed  front form a absorbing circle of attractive propagation. In Figure 15, at time step 130dt, the front of electric wave, Ex inside the GLLH EM cloak, 
$R_i \le r \le R_o$, 
propagates slower than speed of light. The left front propagates outward arrive to left boundary without time delay, The dashed points on the right rear and left to right front form an absorbing circle of attractive propagation. Figure 16 shows that at time step 133dt, the front of electric wave, Ex inside the GLLH EM cloak, , propagation slower than speed of light. The left front propagates outward arrive left boundary with time delay, The dashed points on the right rear and left to right  front form an absorbing circle of attractive to inner circle.  In Figure 17, The red source S located left, at time step 119dt, the front of electric wave, Ex inside the GLLH EM cloak , 
$R_i \le r \le R_o$, 
propagates slower than speed of light. Upper downward break mouth and lower upward break mouth closed, the created wave like a crescent . The dashed points left denote discontinuous absorbing propagation, discontinuous absorbing propagation, right wave front  is created wave, discontinuously. In Figure 18, The red source S located botton, at time step 119dt, the front of electric wave, Ex inside the GLLH EM cloak, 
$R_i \le r \le R_o$, propagates slower than speed of light. Left rightward break mouth and right  leftward break mouth closed, the created wave like a crescent. The dashed points botton denote discontinuous absorbing  propagation, up wave front  is created wave, discontinuously. Figure 19 shows that at time step 84dt,  shows the front of electric wave, Ex inside the GLLH EM cloak, 
$R_i \le r \le R_o$, 
propagates slower than speed of light, the dashed points right behind denote discontinuous absorbing propagation. This indicates  the wave propagation in GLLH cloak without superluminal problem. The red source S located on the right. Figure 20 shows that at time step 84dt, shows the front of electric wave, Ex  inside of the Pendry cloak, 
$R_i \le r \le R_o$. 
When wave arrives at r=Ri,  the front forms closed circle around the inner boundary . This indicates, on inner circle r=Ri, the wave propagation in Pendry cloak has infinite speed. The red source S located on the right. In Figure 21, at time step 90dt, shows the front of electric wave, Ex  inside the GLLH EM cloak , 
$R_i \le r \le R_o$,
propagates slower than speed of  light. The dashed points right behind denote discontinuous absorbing propagation. This indicates the wave propagation in GLLH cloak without superluminal problem. The red source S is located on the right. In Figure 22, at time step 90dt, shows the front of electric wave, Ex inside the Pendry cloak 
$R_i \le r \le R_o$,
where propagation exceeds the speed of light . The red source S is located on the right. The electromagnetic wave propagation in Pendry cloak has superluminal propagation.

\section { INVISIBLE PROPERTIES OF GLLH EM CLOAK}
By using the GL EM Metro Carlo inversion method in [ 5 ], we solve the GL EM cloak inversion $ ( 11)-(15)$ with a distinctive class of materials $a_{\alpha \beta}\log ^\alpha (b_{\alpha \beta}/h) h^\beta$ [3][4],   and obtain the GLLH EM invisible cloak. The invisibility properties of the GLLH EM cloak are verified in this section. 

\subsection { The Invisibility of The GLEM Cloak}
${\it \bf Statement \ 1}$ The exterior EM field is not scattering interfered by the cloak

Because of our GLLH EM exterior cloak inversion, by substituting the GLLH EM cloak parameters in ( 2)–( 5 ) and the EM wave field by GL modeling [ 3 ] into ( 11 ) and ( 12 ), such that integral equations ( 11 ) and ( 12 ) hold —we have identity ( 6 ): no reflection and no scattering from the cloak to interfere the exterior EM field in free space. 

By calculations, when   $R_i=0.2491$ meter and, $R_o =0.47$  meter, we prove that on the outer boundary of the cloak  $r=R_o$,  and   in $ (2) – (5)$ satisfy the condition for any exterior source and frequency. Therefore, any exterior EM wave field must not be scattering interfered by the cloak. By the 3D GL EM modeling [ 3 ] simulations, electric wave propagations in $Figures 5–18$ verify that there is no reflection and no scattering from the cloak to interfere with the exterior EM field in free space. 

By calculations, when $R_i=0.2491$ meter and $R_o =0.47$ meter, we can prove that on the outer boundary of the cloak $r=R_o$,
$\bar \varepsilon$ and $\bar \mu  $ in (2)-(5)
satisfy 
\begin{equation}
\begin{array}{l}
 \varepsilon _r  = \varepsilon _\theta   = \varepsilon _\phi   = 1,\,\,\,r \in R_o , \\ 
 \mu _r  = \mu _\theta   = \mu _\phi   = 1,\,\,\,r \in R_o . \\ 
 \end{array}
\end{equation}
This is a condition for the invisibility of the cloak. From the GLLH EM cloak exterior inversion equations ( 11 ) and ( 12). The condition (17) can be relaxed to  
$n(R_0) = \sqrt{\varepsilon_r \mu_\theta(R_0)} = 1$ in (20). The spherical annular, $R_i \le r \le R_o$ can be arbitray annular. In this simulation,  $R_i=0.2491$ meter and $R_o =0.47$  meter.
 \hfill\break
\hfill\break 
${\it \bf Statement \ 2}$ Any exterior EM wave field can not 
penetrate into the concealment.

From the GLLH EM inner cloak inversion (13)
and (14), and EM integral equation (8) and (9),
the statement 2 is proved.
Also, because on the inner boundary $r=R_i$,the refractive index 
$\sqrt{\varepsilon _r \mu _\theta} $ or
$\sqrt{\varepsilon _\theta \mu _r}$ in (4)-(5) 
is infinite, from the convergence of integral (11)-(14), requires that
the EM wave field  must be vanished on the inner
boundary $r=R_i$. We obtain the condition
\begin{equation}
\begin{array}{l}
 E_\theta   = E_\phi   = 0,\,\,r \in R_i  \\ 
 H_\theta   = H_\phi   = 0,\,\,r \in R_i  \\ 
 \end{array}
\end{equation}
Conversely,  the conditions (17) and (18) are
sufficient condition for the invisibility of the GLLH EM cloak.

\subsection {Without Exceed Light Speed Violation}

${\it \bf Statement \ 3}$ The refractive index in the GLLH EM cloak
 in this paper is large than one, the  phase speed in the cloak is finite and less than light speed.
\hfill \break
 \break
From (4), the $\varepsilon _r  $ and $\mu_r$ are finite positive and monotone
increase function on variable $r$ in the $R_i \le r \le R_o$.
moreover, their minimum on the inner boundary of the cloak are
\begin{equation}
\varepsilon _r (R_i ) = \mu _r (R_i ) = \frac{{2R_o^2 (R_o  - R_i )}}{{R_i^2 (1 + 4R_o  + 2R_o^2 )}}> 0,
\end{equation}
their maximum value on the outer boundary $R_o=0.47$ of the cloak are one.

From (5), the $\varepsilon _\theta  $ and $\mu_\theta $ are monotone
decrease function on variable $r$ in the $R_i \le r \le R_o$,
moreover, their maximum value on the inner boundary of the cloak are infinite, their minimum value on the outer boundary, $R_o=0.47$ meter of the cloak are one

The refractive index of the GLLH EM cloak in (2)-(5) is,
\begin{equation}
n(r) = \sqrt {\varepsilon _r \mu _\theta  }  = \sqrt {\varepsilon _\theta  \mu _r }  \ge 1.
\end{equation}
The phase speed is finite and less than light speed.

Therefore, in the annular configuration $R_i \le r \le R_o$, with $R_i=0.24$ and $R_o=0.47$ 
The GLLH EM invisible cloak material in (2)-(5)  has no violation of exceeding light speed.  The novel electric intensity wave
propagations, which are shown in Figure 3 -18,  in the GLLH EM cloak in (2)-(4) verify that the GLLH EM cloak 
is invisible cloak without exceed light speed violation.

\subsection{Wave Front Branching}

The EM wave propagation in the GLLH EM cloak is completely novel, as displayed in Figure 1 and in Figures 3 to 17. In particular, in Figure 12, the EM wave front is curved around the concealment, and its upside and downside parts intersect at a movable branching point, forming a 2D subsurface. Figure 13 shows that the two wave fronts propagate separately from the branching point. The first wave front inside of cloak is created wave and forward propagating outgoing. The second wave front is attractively propagating and approaching to the inner boundary $r=R_i$ . The speeds of both wave front propagations slower than the light speed. The second wave front attractive propagation is very slow. Its amplitude rapidly decays to zero. Its propagation speed approaches zero. The front branching points form a movable 2D subsurface that depends on the source location. The front branching point and source are separately located on opposite sides of the concealment region. In Figures 12 and 13, the source ( denoted by red S ) is on the right side of the cloak, while the front branching point is on  the left side of the concealment. In Figure 17, the source ( red S ) is on the left side of the cloak, and the front branching point is  on the right side of the concealment. In Figure 18, the source (red S) is located at bottom of the cloak, and the front branching point is located in at the top side of the concealment. These novel EM wave propagations in the GLLH EM cloak and the wavefront branching phenomenon are displayed only in this paper. They do not appear in other cloak papers. The GLLH EM cloak, novel wave propagations, and wavefront branching are patented and copyrighted by the authors at GL Geophysical Laboratory. 

${\it \bf Statement \ 4}$ The concealment region of the GLLH invisible cloak, with the basic EM materials $\varepsilon_0 $ and $\mu_0$ or normal material with refractive index  $n \ge 1$, then no external incident wave can not penetrate the concealment.
Statement 4 theoretically confirms that our discovered GILD cloak phenomenon in 2001 in [ 7 ], display in Figure 1 and Figure 2 .

${\it \bf Statement \ 5}$ When the source is located outside the GLLH invisible cloak in ( 2 )-(5 ), the observer at r is inside the cloak and approaches the inner boundary  $r=R_i$ , the EM wave fields decay to zero in the inverse radial direction. We have 
\begin{equation}
\begin{array}{l}
 E_r (r,\theta ,\phi ) = E_\theta  (r,\theta ,\phi ) = E_\phi  = 0,when \ r = R_i , \\ 
 H_r (r,\theta ,\phi ) = H_\theta  (r,\theta ,\phi ) = H_\phi = 0,when \ r = R_i , \\ 
 \end{array}
\end{equation}
where  $\ r = R_i $ is the inner boundary of the GLLH cloak,  on the inner boundary $\ r = R_i $,  the radial electric wave field $ E_r (r,\theta ,\phi ) =0$ and the radial magnetic wave field $ H_r (r,\theta ,\phi ) =0$ are distictive property of  the GLLH invisible cloak.  On the inner boundary of other transformed cloak, the radial  wave EM field are not zero.
These novel EM wave propagations in our GLLH EM cloak  and
wave front branching phnomena only display in our paper. They
never appear in other cloaks and never appear in other materials in other papers. 
\section{ GL sphere hamonic modeling and inversion}
\subsection{The GLLH cloak material parameters are radial dependent as follows: }

\[
\begin{array}{l}
 \varepsilon _r  = \mu _r  = \frac{1}{{1 + 4R_o  + 2R_o^2 }}\left( {\frac{{2R_o^2 \left( {R_o  - R_i } \right)}}{{r^2 }}} \right. \\ 
  + \frac{{2 + 6R_o }}{{\log \left( {e^{1/R_o } \left( {R_0  - R_i } \right)/\left( {r - R_i } \right)} \right)}}\frac{{\left( {r - R_i } \right)}}{{R_o r^2 }} \\ 
 \left. { - \frac{{1 + 2R_o }}{{\log ^2 \left( {e^{1/R_o } \left( {R_o  - R_i } \right)/\left( {r - R_i } \right)} \right)}}\frac{{\left( {r - R_1 } \right)}}{{R_o^2 r^2 }}} \right), \\ (4)
 \end{array}
\]
\[
\begin{array}{l}
 \varepsilon _\theta   = \varepsilon _\phi   = \mu _\theta   = \mu _\phi   \\ 
  = \frac{R_o}{{2\log \left( {e^{1/R_o } \left( {R_0  - R_i } \right)/\left( {r - R_i } \right)} \right)\left( {r - R_i } \right)}} \\ 
  + \frac{1}{{2R_o\log ^3 \left( {e^{1/R_o } \left( {R_0  - R_i } \right)/\left( {r - R_i } \right)} \right)\left( {r - R_i } \right)}} \\ (5)
 \end{array}
\]

where the $R_i  \le r \le R_o$ is the spherical annular cloak 
domain,  $R_i$ is inner radius of the annular cloak,
$R_o$ is outer radius of the annular cloak, 
${\bar \varepsilon }$ is the dielectric parameter matrix,  ${\bar \mu }$ is the permeability  parameter matrix, 
$\varepsilon _0 $ is the basic dielectric parameter, $\mu _0 $ is the basic magnetic permeability parameter, ${\varepsilon _r }$ is the relative radial dielectric permittivity, ${\varepsilon _\theta  }$ and ${\varepsilon _\phi  }$  are the relative angular dielectric permittivity,
${\mu _r }$ is relative radial permeability, ${\mu _\theta  }$ and ${\mu _\phi  }$ are relative angular permeability.
 
\subsection{GL sphere harmonic radial equation modeling and inversion}

By the second order radial electric field equation (6) in paper arXiv:1701.02583, the GL sphere harmonic radial equation  is       

\begin{equation}
\frac{\partial}{\partial r} \frac{1}{\varepsilon_\theta} \frac{\partial E_l}{\partial r} - \frac{1}{\varepsilon_r r^2} l(l+1) E_l + k^2 \mu_\theta E_l = 0,
\end{equation}
the $l$ is harmonic component. GL Electric field is
\begin{equation}
E(r, \theta, \phi) = E_r r^2 \mu_r = \sum_{l=0}^{M} E_l(r) \sum_{m=-l}^{l} Y_m^l(\theta, \phi) Y_m^{l*}(\theta_s, \phi_s),
\end{equation}
$Y_m^l(\theta, \phi)$ is the sphere harmonic functions.
For  $l = 1,2,.  .N$, using GL modeling and inversion methods to solve the GL  sphere harmonic radial equarion (22) and continuous boundary conditions on the boundary $t=R_o$ with incident wave, we obtain the GL radial electric field  . $E_i (r)$.  By (23),  $E(r, \theta, \phi)$
and the radial electric tensity field  $E_l (r, \theta, \phi)$ are obtained. Similarly calculation, we can obtain the GL radial magnetic field  $H(r, \theta, \phi)$ and radial  magnetic tensity  field  $H_l (r, \theta, \phi)$ . Then the spherical coordinate is translated to rectangle coordinate,  $E_x(x, y, z)$ and $H_x (x, y, z)$
, etc. are obtained.
The simulation of the electric wave propagation in GLLH invisible cloak is described in section 4, simlarly, we obtained the magnetic wave propagation through the GLLH invisible cloak without superluminal and without time delay.  The method of developping GLLH cloak is GL modeling and inversion with search class $a_{\alpha \beta}\log ^\alpha (b_{\alpha \beta}/h) h^\beta$ that is different from the coordinate transformation. 

\section{ Advanced GLLH Invisibility AGLLH Cloak }
From
\[
\begin{split}
\varepsilon_\theta = \varepsilon_\phi \\
 = \frac{dR(r)}{dr} = {} & \frac{R_o}{2} \frac{1}{\log\left(e^{\frac{1}{R_o}} \frac{R_o - R_i}{r - R_i}\right)} \\
& \times \frac{1}{(r - R_i)} \   \   \   (5) \\
& + \frac{1}{2R_o} \left(\log\left(e^{\frac{1}{R_o}} \frac{R_o - R_i}{r - R_i}\right)\right)^{-3} \frac{1}{r - R_i},
\end{split}
\]

We find an original fuction

\begin{equation}
\begin{split}
R(r) = {} & -\frac{R_o}{2} \log\left(\log\left(e^{\frac{1}{R_o}} \frac{R_o - R_i}{r - R_i}\right)\right) \\
& + \frac{1}{4R_o} \left(\log\left(e^{\frac{1}{R_o}} \frac{R_o - R_i}{r - R_i}\right)\right)^{-2} \\
& - \frac{R_o}{2} \log(R_o) - \frac{5}{4} R_o
\end{split}
\end{equation}
The quasi- topological transformation (24)  maps the spherical annular region, $R_i \le r \le R_o$
  in the positive space one to one onto the exterior of the negative sphere  in the negative space,$-\infty \le R \le -R_o$
 .which  induces an advanced anisotropic electromagnetic invisibility AGLLH cloak (5). Unlike the topological mapping in[25], the mapping (24) is a quasi-topological transformation.
The radial magnetic wave propagation through the AGLLH cloak and GLLH cloak are bounded and displayed in the below figures. In Figure 23, at at time step 90dt, shows the front of electric wave, Ex  inside the AGLLH EM cloak
$R_i \le r \le R_o$ propagation, very similarly, In  Figure  21, at at time step 90dt, shows the front of electric wave, Ex  inside the GLLH EM cloak $R_i \le r \le R_o$ propagation.
\begin{figure}[h]
\centering
\includegraphics[width=0.9\linewidth,draft=false]{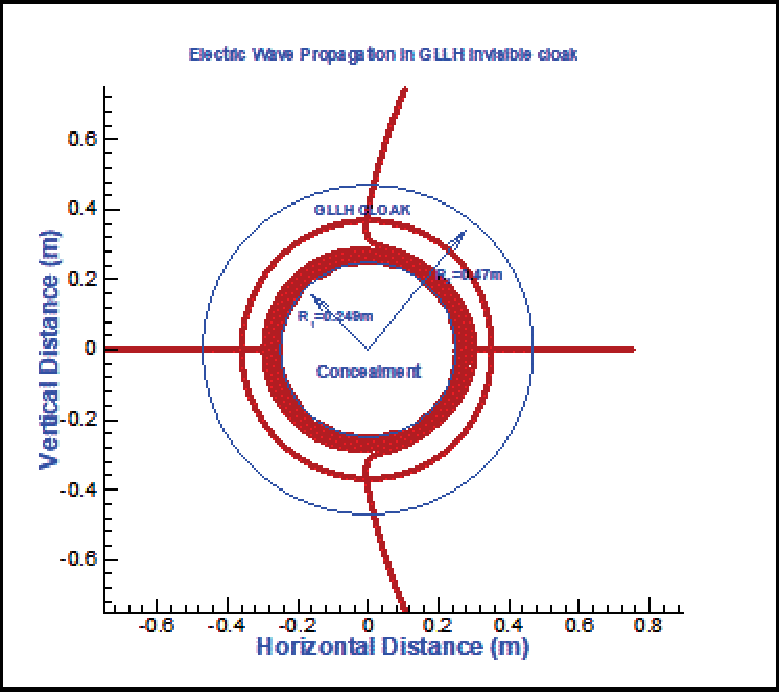}
\caption{ (color online) 
Analytical Electric wave $E_x$ propagation through AGLLH invisible cloak  at time step $90dt$, the two peaks of
the crescent like wave front intersects at a branching point, red S in right is source location}\label{Fig23}
\end{figure}
\section{ Discussion and Conclusion}
\subsection {Discussion} 
We discovered the GILD clock phenomenon to prevent detection by external EM waves in 2001 [7]. Repeated GILD simulations show that the GILD cloak is a mathematical-physical phenomenon. In 2003, we developed a completely new GL EM modeling [ 3 ] and GL EM inversion [ 5] to investigate this strange phenomenon. We found that seeking a proof of the Goldbach conjecture and finding a completely invisible cloak with a refractive index greater than 1 are both non scattering inverse problems[8]. Congruence is characterized by waves, while the distribution of prime numbers acts like a refractive index. Chen Jingyun’s GOLDBACH $P_x(1,2)$   published on Chinese stamps and published in People's Daily, it became a household name and formula.
\begin{figure}[h]
\centering
\includegraphics[width=0.9\linewidth,draft=false]{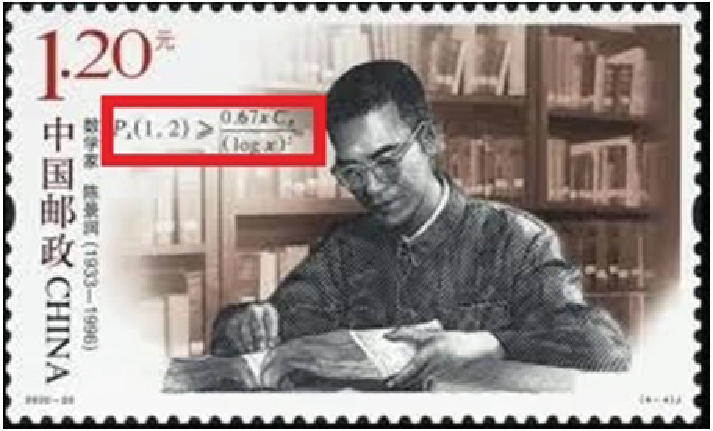}
\caption{ (color online) 
Chen Jingyun’s GOLDBACH $P_x(1,2)$   published on Chinese stamps and published in People's Daily,  it became a household name and formula.}\label{Fig24}
\end{figure}

Anyone chinese who encounters bottlenecks and difficulties will think of him. From Chen Jingyun’s GOLDBACH proof clue.  Anyone chinese who encounters bottlenecks and difficulties will think of him. From Chen Jingyun’s GOLDBACH proof clue, $P_x(1,2) \ge \frac{0.67 x C_x}{(\log x)^2}$ [4], while the distribution of prime numbers, $\frac{x C_x}{(\log x)^2}$, acts like a refractive index of the congruence characterized waves, so we choose,${\it a_{\alpha \beta}\log ^\alpha  (b_{\alpha \beta}/h) h^\beta}$, $h=r-R_i$ as search material class for GL non scattering modeling and inversion[3][5]. We obtained GLLH invisible cloak and its relative radial dielectric permittivity and magnetic permeability,
\[
\begin{array}{l}
 \varepsilon _r  = \mu _r  = \frac{1}{{1 + 4R_o  + 2R_o^2 }}\left( {\frac{{2R_o^2 \left( {R_o  - R_i } \right)}}{{r^2 }}} \right. \\ 
  + \frac{{2 + 6R_o }}{{\log \left( {e^{1/R_o } \left( {R_0  - R_i } \right)/\left( {r - R_i } \right)} \right)}}\frac{{\left( {r - R_i } \right)}}{{R_o r^2 }} \\ 
 \left. { - \frac{{1 + 2R_o }}{{\log ^2 \left( {e^{1/R_o } \left( {R_o  - R_i } \right)/\left( {r - R_i } \right)} \right)}}\frac{{\left( {r - R_1 } \right)}}{{R_o^2 r^2 }}} \right), \\ (4)
 \end{array}
\]
and its relative angular dielectric permitivity and magnetic permeability,           
\[
\begin{array}{l}
 \varepsilon _\theta   = \varepsilon _\phi   = \mu _\theta   = \mu _\phi   \\ 
  = \frac{1}{{2\log \left( {e^{1/R_o } \left( {R_0  - R_i } \right)/\left( {r - R_i } \right)} \right)\left( {r - R_i } \right)}} \\ 
  + \frac{1}{{2\log ^3 \left( {e^{1/R_o } \left( {R_0  - R_i } \right)/\left( {r - R_i } \right)} \right)\left( {r - R_i } \right)}} \\ (5)
 \end{array}
\]
\begin{equation}
\varepsilon_r(R_o) = \mu_r(R_o) = \frac{(R_o - R_i)}{R_o^2},
\end{equation}
\begin{equation}
\varepsilon_\theta(R_o) = \varepsilon_\phi(R_o) = \mu_\theta(R_o) = \mu_\phi(R_o) = \frac{R_o^2}{(R_o - R_i)}
\end{equation}

\begin{equation}
\begin{split}
n^2(r) =\\
 {} & \mu_\theta \varepsilon_r = \varepsilon_\theta \mu_r = \\
& \frac{1}{1+4R_o+2R_o^2} \Big( R_o^3 \frac{R_o - R_i}{r^2} \\
\frac{1}{\log\left(e^{\frac{1}{R_o}}
 \frac{R_o - R_i}{r - R_i}\right)(r - R_i)} \\
& +  R_o \frac{R_o - R_i}{r^2} \\
\frac{1}{\log^3\left(e^{\frac{1}{R_o}} \frac{R_o - R_i}{r - R_i}\right)(r - R_i)} \\
& + \frac{2+6R_o}{2\log^2\left(e^{\frac{1}{R_o}} \frac{R_o - R_i}{r - R_i}\right)r^2} \\
- \frac{1+2R_o}{2R_o\log^3\left(e^{\frac{1}{R_o}} \frac{R_o - R_i}{r - R_i}\right)r^2} \\
& + \frac{2+6R_o}{2R_o^2\log^4\left(e^{\frac{1}{R_o}} \frac{R_o - R_i}{r - R_i}\right)r^2} \\
- \frac{1+2R_o}{2R_o^3\log^5\left(e^{\frac{1}{R_o}} \frac{R_o - R_i}{r - R_i}\right)r^2} \Big),
\end{split}
\end{equation}
\begin{equation}
n(R_o) = \sqrt{\varepsilon_r(R_o)\mu_\theta(R_o)} = \sqrt{\mu_r(R_o)\varepsilon_\theta(R_o)} = 1,
\end{equation}
\begin{equation}
n(r) = \sqrt{\varepsilon_r \mu_\theta} = \sqrt{\varepsilon_\theta \mu_r} \ge 1.\   \   \  \  (20)
\end{equation}
The estimates in [21]–[23] are useful for invisible inversion. By calculations, when  $R_i =0.2491$ meter and $R_o =0.47 $ meter, we can prove that on the outer boundary of the cloak on $R_o =0.47 $  , relative $\bar{E}$ and  $\bar{M}$  in ( 2)–( 5 ) satisfy
\[
\varepsilon_r = \varepsilon_{\theta} = \varepsilon_{\phi} = 1,
\mu_r = \mu_{\theta} = \mu_{\phi} = 1,  (17) \]
The GL method [3][5] combines analytical harmonic method and numerical method. It can perform approaches, enabling  analytical, numerical, or mixed field simulation and material generation. Discontinuous wave front branching is a completely novel propagation properties. 
Note that the formula (5) is a correction of the typo of the formula (5) in 2010 in $arXiv:1005.3999$ in [1]. When  $R_o =1.0 m$ , both are same. The typo has been corrected since 2011. 
By comparison, in Figure 19, at the 84th time step, the EM wave propagates in the GLLH EM cloak more slowly than the speed of light. In Figure 20, at the 84th time step, the EM wave propagates through the Pendry cloak with infinite speed, and its front coincides with the inner sphere. Figure 22 shows that the EM wave inside the Pendry cloak [14] exhibits superluminal propagation. At the 90th time step, EM propagation in the GLLH EM cloak is presented in Figure 21; it is obvious that the speed of the EM wave inside the GLLH EM cloak does not exceed the speed of light. To overcome the infinite speed in [ 14 ], Ulf Leonhardt et al. proposed a new cloak with finite speed based ,on a Euclidean and non - Euclidean joint transform [ 17 ]. The ULF cloak overcomes the " infinite speed " physical violation, even though its refractive index is less than one with superlumial in some sub domains ,$n(\theta, \theta', r) = n(0.75\pi, \pi, r) < 1$. H. Chen, B. Wu, B. Zhang, A. Kong provide a proof [24] for [14]. 

\subsection {CONCLUSION }
There are two kinds of inversions: scattering inversion [18] and non-scattering inversion [3][5]. Theoretical and GL spherical harmanic numerical analysis of GL EM modeling and non-scattering inversion , along with numerous GLLH EM modeling simulations, verified that the GLLH EM cloak has full invisibility functions. The GLLH EM invisible cloaks avoid the two violations of infinite speed and superluminal propagation, and avoid the time delay. The GLLH invisible cloak is not from coordinate transformation.  
The novel EM wave propagations in GLLH invisible cloak presented in Figure 1 in 2010[1] closely match the GILD cloak phenomena in Figure 2, which were discovered in 2001[7].
In Figure 1 and Figures  $3 - 16$. The front inside the GLLH EM cloak discontinuously around the concealment, forming a crescent - like shape in Figure 1 and Figure 12. The two front teeth of the crescent front intersect at the branching point . The branching points form a 2D subsurface. In Figure 13  and 14, the wave front splits into two fronts at the front branching point. The outgoing front propagates forward and recovers to original wave front in free space. The attractive front propagates and shrinks to the inner boundary of the cloak, with its amplitude and speed rapidly decaying to zero. The incident waves entering the GLLH invisible cloak are all absorbed by the materials; simultaneously, created waves propagate outward, such that electromagnetic waves propagate through the GLLH invisible cloak without superluminal speeds and without time delay. The GLLH invisible cloak opened a non-scattering GL modeling and inversion approach to create an invisible cloak without superluminal effects and without time delay. From the GLLH invisible cloak, we discovered superscience, positive space and negative space, and invisible science; the isotropic electromagnetic invisible GLHUA sphere; and the anisotropic electromagnetic invisible GLHUA cloak. After many years of progress[12][13][15][16] by GL modeling and inversions, the complex relative angular permittivity and permeability in (5) of the GLLH cloak in 2010  is simplified to $\varepsilon_{\theta} = \mu_{\theta} = \frac{(R_o - R_i)^2}{(r - R_i)^2}$  in GLHUA invisible cloak[10] in 2016.

A point in space has no length or distance in any direction; any movement passing through a point is instantaneous, and the time required is zero. The linear coordinate transformation   $r = R_i + \frac{(R_o - R_i)R}{R_o}$ in [14] maps the point  $R=0$ to a spherical surface  $r=R_i$ . The time required for the wave to pass the surface $r=R_i$ is equal to the time required for the same wave to pass the point $R=0$  , so the time required for the wave to pass the surface $r=R_i$  is zero, which induces an infinite phase velocity for the wave passing the surface  $r=R_i$. The linear coordinate transformation  $r = R_i + \frac{(R_o - R_i)R}{R_o}$  in [14] induces the relative refractive index and superlumial propagation. In section 7, we proposed quasi topological coordinate transformation,
\begin{equation}
\begin{split}
R(r) = {} & -\frac{R_o}{2} \log\left(\log\left(e^{\frac{1}{R_o}} \frac{R_o - R_i}{r - R_i}\right)\right) \\
& + \frac{1}{4R_o} \left(\log\left(e^{\frac{1}{R_o}} \frac{R_o - R_i}{r - R_i}\right)\right)^{-2} \\
& - \frac{R_o}{2} \log(R_o) - \frac{5}{4} R_o \ \ \ (24)
\end{split}
\end{equation}

The quasi topological transformation (24)  maps the spherical annular region $ R_i \le r \le R_o$   in the positive space one to one onto the exterior of the negative sphere  in the negative space,$-\infty \le R \le -R_o$ , maps surface $r=R_i$ in positive space to negative surface in negative space $-R = -R_n \rightarrow -\infty$ , which  induces an advanced anisotropic electromagnetic invisibility AGLLH cloak with relative  angular permittivity and permeability  (5) and relative refractive index (27). The GLLH and AGLLH EM invisibility cloaks with the relative refractive index  overcome the fundamental difficulty caused by mapping a point to a surface in positive space via coordinate transformation. The electromagnetic environment within the concealment zone of a single-layer cloak has been discussed [19]. Authors Feng Xie and Lee Xie proposed a new mirage [20]. The GLLH EM cloak, GLLH EM cloak modeling and inversion, and GLLH EM cloak software are patented and copyrighted by authors in GL Geophysical Laboratory. 

\subsection{Open Question}
Open question: Can we construct a 3D Kakeya set [26[[27]where line segments of length   represent the vector field of electromagnetic wave ray-tracing propagation through GLLH[1], GLHUA cloaks[10], and the GLHUA sphere[9]?

\subsection{Acknowledgments}
Authors thank Professor P.D.Lax, Professor Y.M.chen, Professor Jiaqi Liu,Professor Feng Kang, Professor Lin Qun,Professor Micheal Oristaglio, Professor Chen Jingyun, Professor Guo Youzhong, Professor He Zi Fan , Professor Zhu Yanbing, Professor Enrest Majer, Professor Erica Jen,Professor R. P. Serivastav, Professor W.J.Kim, Professor S.S.Chern, Professor C.N.Yang, Professor KiHa Lee for their help. Author Ganquan Xie thanks Professor Yongen Cai for his invitation to speak at the College of Earth and Space Sciences at Peking University on May 9, 2008. The title of the presentation was 'Global and Local Field Modeling and Inversion.' During the talk, Xie forewarned that a major disaster would happen on Chinese land, which was confirmed by the Wenchuan earthquake three days later on May 12, 2008. During the subsequent Wenchuan earthquake investigation, Xie discovered a seismic cloak protecting Chengdu, which inspired the development of the GLLH invisibility cloak. Author Ganquan Xie thanks to the leaders of Wangcheng District, Changsha City, Secretary Tan Xiaoping, Director Yang Dongfang, Director Hu Zhenfu, Director Yao Wangshu, Secretary Wu Nanjun, Secretary Mao Bin, Secretary Xie Zhixiong, and leaders of the Science and Technology Bureau, Director Longjun Peng Yang Director Wenxia Secretary and Zhou Xiangqi, Li Xiaoqing Section Chief Engineer,Director Liu Zhiqiang, Secretary Wu Sheng, Professor Deng Qihua,Professor an Shunshi etc for their help. Author Ganquan Xie thanks to Principal Hu Ailong, principal of Wangcheng No. 2 Middle School, principal Liang Zhijun, principal Chen Manfu, principal Sun Yong, principal Li Hongjun,princial Chen Kexin, princial Zhou Ming, princial Liu Jianjun and principal Fu Liu Jiansheng, and Hunan Supercomputing Science Society Professor Luo Jianshu, Professor Xie Xiaoliang, Professor Tian Zhongchu, Professor Cao Shigu, Professor Tan Shusheng, Professor Tang Jingtian, Professor Zhuang Hao, Engineer Huang Hu, Director Ouyang Junli, Professor Zho BiHua, Professor ShuFang Yang, Professor Enginner Li Jing, Professor Zide Yang, Engineer Huiyi Yu.Professor Cai Zhaogui,Professor Xie Xinsheng, Professor Zhu Daxin, Engineer Zhu Jianpeng, Professor Zhuo Xianwei, Professor Li Youming, Professor Liu Hong etc for their help.

\subsection{Explanation}
Author Contribution Declaration, Corresponding author Jianhua Li is the research leader of this paper and our invisibility science, ideas, concepts, formulas, and overall organization and arrangement. Professor Feng Xie, Lee Xie, and Ganquan Xie contributed to the ideas, concepts and formulas of this paper. Feng Xie wrote the software; Lee Xie made the figures and videos.

Funding declaration, There is no funding to support our paper “GLLH EM Invisible Cloak With Novel Front Branching And Without Exceed Light Speed Violation”. We have disabilities and volunteered for invisibility science for 26  years, never applying any funding to support our research “Electromagnetic invisible GLLH cloak” and “Electromagnetic invisible GLLH Sphere” and no funding for our “invisible science” research.

The data availability statement,
All data generated or analyzed during this study are included in this published article.
All invisible GLLH cloak analyzed during this study are included in this published article by
corresponding author Jianhua Li and co authors. In this paper and our study, corresponding author Jianhua Li and all co authors did not use any external data. All of the material and contents in this paper is owned by the authors in this paper. Fortran programs for the simulations in this paper were written by us.



\end{document}